\documentclass[12pt]{article}
\usepackage{amsmath}
\usepackage{amssymb}
\usepackage{graphicx}
\usepackage{color} 
\usepackage{chemarrow}
\usepackage{subfigure}

\usepackage[round,sort&compress,comma,authoryear]{natbib}
\newcommand*{\doi}[1]{\href{http://dx.doi.org/#1}{doi: \nolinkurl{#1}}}

\usepackage{setspace} 
\usepackage[labelfont=bf,labelsep=period,justification=raggedright]{caption}

\date{}
\begin{document}
\title{City traffic forecasting using taxi GPS data: \\
A coarse-grained cellular automata model}
\author{Yucheng Hu$^{1*}$, Minwei Li$^{2}$, Hao Liu$^{2}$, Xiaolu Guo$^{3}$, \\
Xiaowei Wang$^{3}$ and Tiejun Li$^{3}$}
\maketitle

\begin{flushleft}
1. Zhou Pei-yuan center for applied mathematics, 
Tsinghua University, Beijing, China, 100008
\\
2. Beijing Transportation Information Center, Beijing Key Laboratory of Integrated Traffic Operation Surveillance and Service, Beijing, China, 100161
\\
3. Laboratory of Mathematics and Applied Mathematics and School of Mathematical Sciences,
Peking University, Beijing, China, 100871
\\
$\ast$ E-mail: Corresponding huyc@tsinghua.edu.cn
\end{flushleft}

\newpage
\begin{abstract}
City traffic is a dynamic system of enormous complexity. Modeling and predicting city traffic flow remains to be a challenge task and the main difficulties are how to specify the supply and demands and how to parameterize the model. In this paper we attempt to solve these problems with the help of large amount of floating car data. We propose a coarse-grained cellular automata model that simulates vehicles moving on uniform grids whose size are much larger compared with the microscopic cellular automata model. The car-car interaction in the microscopic model is replaced by the coupling between vehicles and coarse-grained state variables in our model. To parameterize the model, flux-occupancy relations are fitted from the historical data at every grids, which serve as the coarse-grained fundamental diagrams coupling the occupancy and speed. To evaluate the model, we feed it with the historical travel demands and trajectories obtained from the floating car data and use the model to predict road speed one hour into the future. Numerical results show that our model can capture the traffic flow pattern of the entire city and make reasonable predictions. The current work can be considered a prototype for a model-based forecasting system for city traffic.
\end{abstract}


\section{Introduction}
Traffic congestion has become a serious problem limiting the efficient operating of many big cites. In order to avoid or easy the traffic congestion, traffic prediction is becoming more and more important for trip planing and traffic management. However, due to the complexity of urban traffic, accurate traffic forecasting is extremely difficult. After decades of tremendous efforts, the mathematical theory of traffic flow on single road or simple network is now well-established~\citep{treiber1999derivation, nelson2000synchronized, hoogendoorn2001state}. However, the modeling of large scale urban traffic dynamics has fall far behind. For quite a long time the lack of high quality traffic data has been impeding the development of traffic modeling and prediction on large scale road networks. Recently, large amount of floating car data and mobile phone data covering the whole urban area with high spatial and temporal resolution have become available. These data provide new sources of information, allowing us to push the limit of traffic modeling and prediction on city-wide scale~\citep{gonzalez2008understanding, herrera2010evaluation}.

In this paper we consider the problem of short-term prediction (one hour into the future) of the traffic speed on city-wide scale. Existing traffic prediction approaches can be roughly divided into statistical methods and model-based methods. Statistical approaches usually use time series analysis~\citep{min2011real, clark2003traffic} or machine-learning techniques~\citep{furtlehner2007belief, lv2015traffic, nguyen2008survey} to explain the data and make the prediction. If well implemented, they can achieve quite good accuracy in short-term traffic prediction. However, statistical models are often criticized as ``black boxes'' and they cannot handle situations that are not recorded in historical data.  As opposed to the statistical methods, model-based approaches try to model the traffic flow with certain physical or empirical rules. There are discrete models which focus on the driving behavior of individual cars~\citep{nagel1992cellular, chopard1996cellular, esser1997microscopic}, and continuous models which study the evolution of macroscopic quantities such as the average road speed and occupancy~\citep{lighthill1955kinematic, treiber1999derivation}. There are also hybrid methods that combine the discrete and continuous models in a hierarchical way to deal with traffic systems on large scale~\citep{zhou2014dtalite, li2015hierarchical}. In practice, however, modeling the traffic in a complex urban environment is not easy: one need to specify the detailed road supply and traffic demands and parameterize the model. Fulfilling these tasks usually requires lots of tedious preparation and calibration work, or must use some heuristic assumptions. 


Here we propose a new and relative simple approach for city traffic modeling and simulation. Our method is model-based, but also requires heavily use of statistical tools in setting up the model. Due to the limitation of the computation power and data (we only have the floating data of the taxis, but not all the vehicles), we do not aim to faithfully model the real traffic system in fully detail. Instead, we just want to capture the effective traffic dynamics. In particular, we simulates a large number of vehicles moving on uniform grids. The grid system has a spatial resolution of $100m \times 100m$, which is much large compared with the microscopic cellular automata models. Now that one grid can hold more than one cars, so we call it the coarse-grained cellular automata model. This model is different than continuous traffic models as it keeps track of the individual vehicles. It is also different than the discrete models in that there is no explicit car-car interaction. Instead, vehicles are coupled through the coarse-grained state variables. Because of these differences, we cannot directly import the parameters of other traffic models into our case. 

The main contribution of this work is that we parameterize the model by fitting it to the historical data. For each grid, we learn a coarse-grained fundamental diagram from the flux-occupancy relation estimated from the historical data. From these fundamental diagrams we are able to extract the free-flow speed and road-capacity at each grid (since there is no explicit road in our model, these quantities measure the effective supply of the grids). In addition, the fundamental diagrams also tells us how the traffic speed changes with the occupancy, which is the key information required by our model. In the end, we obtain an effective traffic flow system for the taxis of the entire city. Because taxis share the same road with other vehicles, the simulated speed of this effective system can be used to predict city-wide traffic speed. To evaluate the model, we extract the Origin-Destination (O-D) demands and driving trajectories of taxis from historical floating car data and feed it to the model (for practical traffic forecasting this part will be replace by simulated O-D demands and routes in the future). Promising results are obtained: on the city-wide level, the simulated traffic flow pattern is similar to that observed in the historical data; on the road-segments level, the model predicted traffic speed (one hour into the future) scores a reasonable accuracy. This work can serve as a starting point of a practical city traffic forecasting system, on top of which further improvements can be made in future work. 

The rest of the paper is organized as follows. In Sec.~\ref{sec_preprocess} we consider the data pre-precessing procedures for noise-removal. In Sec.~\ref{sec_estimation} we discuss how to estimate traffic state from data. In Sec.~\ref{subsec_model} the coarse-grained cellular automata model is built. Numerical results and prediction errors are presented in Sec.~\ref{sec_result}. Finally in Sec.~\ref{sec_discuss} we discuss several possible improvement that can be made to the model as topics for future research.

%
%

\section{Data preprocessing}
\label{sec_preprocess}
The floating car data is collected from GPS devices mounted on every taxis in Beijing, China. This kind of device will repeatedly transmit vehicle information to the server, including the current time, GPS coordinate and operating status (0---vacant, 1---occupied, 2---non-operating). The frequency of data transmission varies between 20 seconds to one minutes, depending on the specific standard that the taxi companies use. The total number of taxis in Beijing is about 66,000, continuously generating records each day from 0:00 to 24:00. In total we have data during the period from 2013-3-1 to 2013-3-21, from which only the weekdays days are kept for further analysis (there was no national holiday nor clement weather condition during that period of time).

The raw data contains three types of errors, which are referred as macroscopic, mesoscopic, and microscopic errors depending on their relative size. Macroscopic errors are records whose GPS location is invalid or fall too far away from the city limit. Mesoscopic errors are mainly consist of large spatial displacements ($\ge10$km) between two consecutive time records made by one taxi. Such large displacement is not consistent with any driving behavior, and so should be ruled out. The microscopic errors are due to the limited precision of the GPS devices ($\sim 10$m). In a dense road network, the microscopic errors may cause confusion in matching GPS points to specific roads. 

\begin{figure}
 \centering
 \includegraphics[width=.8\textwidth]{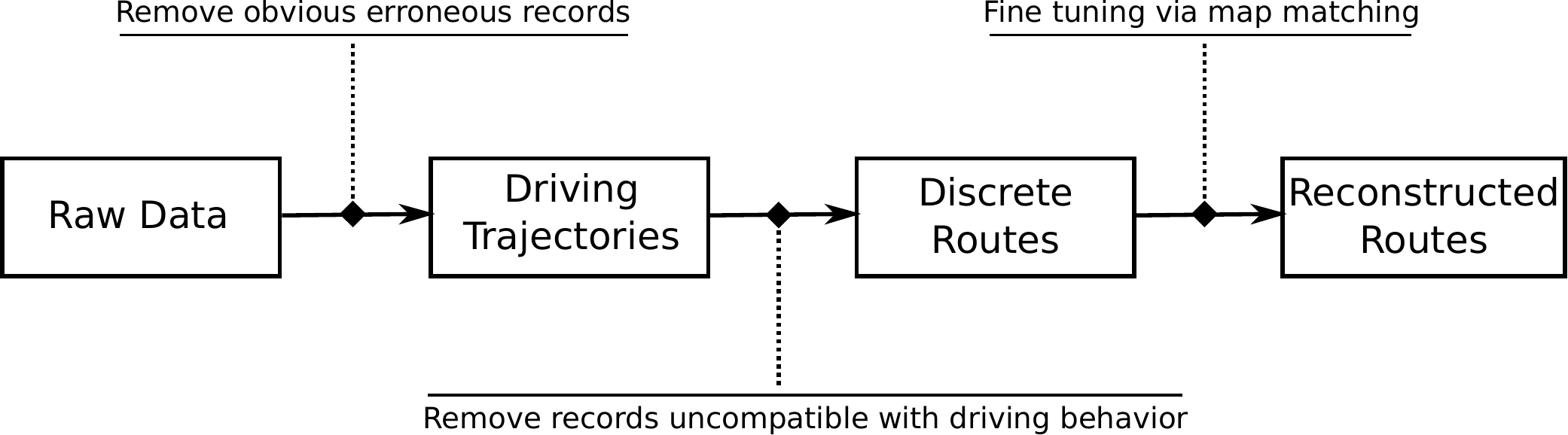}
 \caption{Work-flow of data preprocessing (see main text for details). }
 \label{figDP}
\end{figure}

In order to remove or reduce those errors, we conduct data preprocessing following the pipeline illustrated in Fig.~\ref{figDP}. Firstly, records that are obviously erroneous are removed from the raw data (macroscopic errors). We then group all the records generated by one taxi within one day and sort them by time to generate a \emph{driving trajectory} made by a list of discrete GPS-location points. Considering that the speed of a vacant taxi may not be a good representative of the actual driving speed (for instance it may simply pull over and rest on the side of a road), we only keep records generated by occupied taxis. The vacant/occupied status can be inferred from taxis' operating flag (0 --- vacant; 1 --- occupied) contained in the data. Now for each occupied taxi trajectory, we walk from the first point to the last, remove records that are incompatible with driving behaviors (mesoscopic errors), break up the current route and start a new one when a large time gap is encountered ($\ge 5$mins). By doing so, for each taxi in one day we may obtain multiple \emph{discrete routes}, corresponding to a number of loaded trip this taxi had made in that day. Lastly, we apply a map-matching algorithm provided by an open-source software called OSRM~\citep{luxen-vetter-2011} to map each discrete route to the actual road network. During the process of map-matching the microscopic errors are reduced. In the end we obtain a large set of \emph{continuous routes} reconstructed from all the taxis' floating car data, and from them we can estimate the traffic state such as the average speed and volume.

\section{Traffic state estimation}
\label{sec_estimation}
\begin{figure}
 \centering
 \includegraphics[width=.8\textwidth]{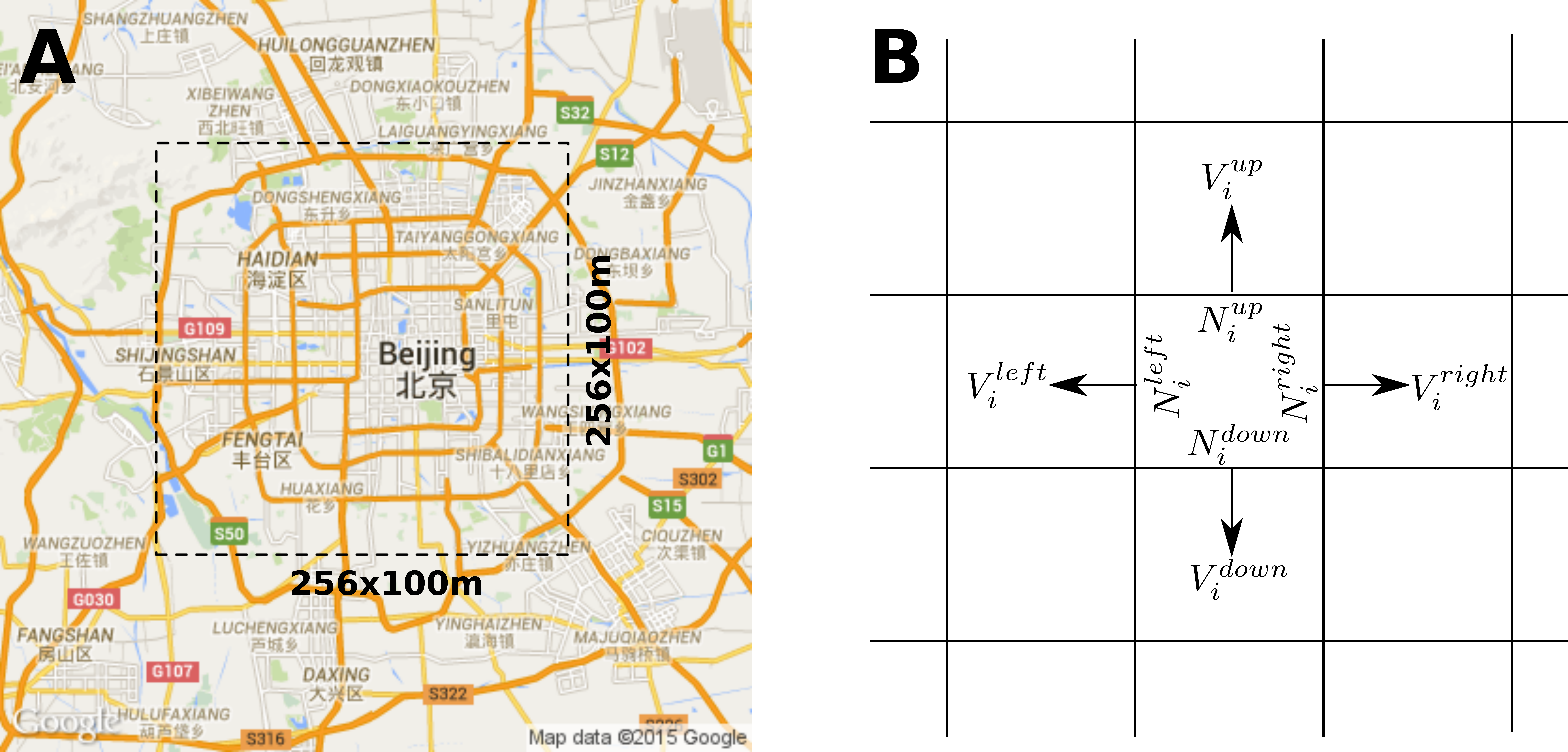}
 \caption{(\textbf{A}) Simulation domain. The $25.6km\times 25.6km$ central part of Beijing is partitioned into grids of $100m \times 100m$ in size, forming a $256\times 256$ uniform grids. (\textbf{B}) Coarse-grained cellular automata model. The traffic state is described by average speed $V_k$ and occupancy $N_k$ for each direction (right, up, left, down) at each grid.}
 \label{figCA}
\end{figure}

In the coarse-grained cellular automata model, the simulation domain (the central Beijing area) is partitioned into $256\times 256$ uniform-grids with $100m \times 100m$ spatial resolution (Fig.~\ref{figCA}\textbf{A}). At each grid, we assume that the cars can move in one of four directions: right, up, left and down ( Fig.~\ref{figCA}\textbf{B}). Denote $V_{i}^{right}$, $V_{i}^{up}$, $V_{i}^{left}$ and $V_{i}^{down}$ as the \emph{average speed} at grid $i$ ($i=1, 2, \cdots, 65536$) in each direction, and $N_{i}^{right}$, $N_{i}^{up}$, $N_{i}^{left}$ and $N_{i}^{down}$ the \emph{average volume} at each grid. In the following we do not include the direction superscript in the notation and simply write $V_{i}$ and $N_{i}$. Unless specified explicitly, operations on them are applied in all the four directions. 

Next we discuss how to estimate $V_{i}$ and $N_{i}$ from historical or simulated \emph{continuous routes}, from which we can deduce a car's position at any given time. First, we define the \emph{instantaneous speed} and \emph{instantaneous occupancy} at time $t$ as
\begin{eqnarray}
 &&\hat{V}_{i}(t) = \sum_j v^j \Lambda^j / \sum_j \Lambda^j, \\
 &&\hat{N}_{i}(t) = \sum_j \Lambda^j, \\ 
 \mathrm{where}  &&\Lambda^j = 
 \begin{cases} 
 1, & \mbox{if car-}j \mbox{ is at grid-} i \mbox{ at time } t,\\ 
 0, & \mbox{otherwise.} 
 \end{cases}
 \label{eq_instspeed}
\end{eqnarray}
The single vehicle speed $v^j$ and the indicator function $\Lambda^j$ can be readily obtained from the reconstructed or simulated continuous routes at any time point $t$. Note that the actual road may be oblique but the road links on the grid system can only be horizontal or vertical. Thus if an oblique road is transformed to the grid system, its zigzagged representation may have a road length larger than the actual road. As a result, the speed we obtained on the grid system may be faster than the real speed. This treatment is equally applied to both the historical data and simulated results and hence will not affect the comparison between them that will be done later in this paper.

One can think of the instantaneous speed and occupancies as one picture frame of the traffic state taken at time $t$. Due to the limited number of probing vehicles ($\sim 30000$ occupied taxis distributed in $256\times256= 65536$ grids), $\hat{V}_i$ and $\hat{N}_i$ computed this way are zeros on many grids. The smallness in sample number also increases the statistical noise. To obtain a more robust estimation, the idea is to take a number of pictures in a short period of time and average the signals. Thus at time $t_n$, the average speed $V_i(t_n)$ is obtained by averaging $\hat{V}_i$ at ten previous time points (including $t_n$), i.~e.,
\begin{eqnarray}
 V_i(t_n) = \frac{1}{10}\sum_{s=0}^9 \hat{V}_i(t_{n-s}).
 \label{eq_averagespeed}
\end{eqnarray}
Similarly, the average occupancy $N_i(t_n)$ is obtained by summing $\hat{N}_i$ at ten previous time points,
\begin{eqnarray}
 N_i(t_n) = \sum_{s=0}^9 \hat{N}_i(t_{n-s}).
 \label{eq_occupancy}
\end{eqnarray}
This kind of data augmentation procedure is based on the assumption that the traffic state does not change too much during the selected time window. In our simulation the time step $t_n - t_{n-1}$ is one minutes, so for the ten-minute time window used here this assumption holds in general. Finally we note that for many grids $V_i$ and $N_i$ are still zeros because many places are seldom visited by taxis. This is totally acceptable and does not affect out further analysis.

\section{Coarse-grained cellular automata model}
\label{subsec_model}
Generally speaking, to build a traffic model, one need to answer three questions: (i) demands: what are the O-D demands and the routes they would take? (ii) supply: what is the road capacity and free-flow speed of each road segment? (iii) driving model: how the cars move on the roads? In this section we will address these questions.

\subsection{Demands}
\label{subsec_demand}
The model simulates an ensemble of cars moving on the grids. To do so we need to know the origin and destination of each trip and the route each individual vehicle takes. For a fully integrated traffic simulation and forecasting system, ideally one need to generate these information on-the-fly. Predicting O-D demands and route choice by itself is a very difficult problem, which will not be discussed in this work. Here however, we only want to give a proof-of-concept test to see if the model can capture the traffic flow dynamics. So instead we extract all the O-D and trajectory information from historical data. The rational behind this is that, with the ``correct" traffic demands as input, the model should be able to regenerate the traffic dynamics observed in the historical data. In this way all the route choices are pre-determined so there will be no dynamical traffic assignment issue.

\subsection{Supply}
\label{subsec_supply}
For simplicity we currently use an implicit road system and the entire urban road network is represented by uniform grids. Each grid contains four effective road segments, pointing to the right, up, left and down direction, respectively. Each road segment may be occupied by multiple cars at the same time. There is no turning restrictions and no traffic lights. From the historical data we are able to estimate the capacity of each road segment may offer: more cars passing means larger road and no car means no road. The detailed procedure of estimating the road capacity will be discussed in Sec.~\ref{subsec_driving}. This grid system can be considered as an approximation to the actual road network.

\subsection{Model}
\label{subsec_driving}
To let the traffic flow, we need to do two things: 
\begin{enumerate}
 \item Given the position of all vehicles on the grids, determine the driving speed on each road segment $V_i$.
 \item Given $V_i$, update the position of all vehicles on road.
\end{enumerate}

\begin{figure}
 \centering
 \includegraphics[width=.8\textwidth]{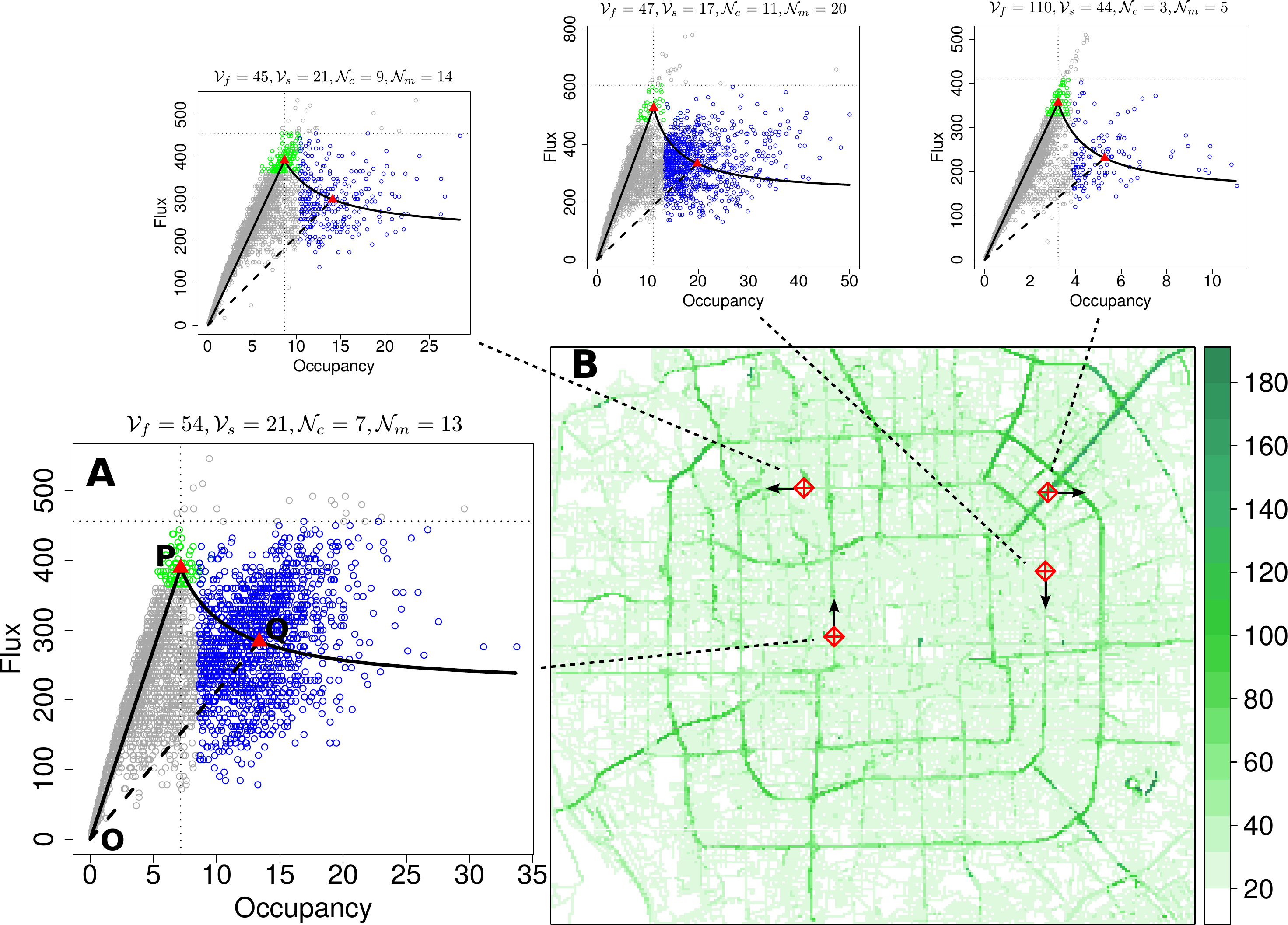}
 \caption{Estimating model parameters from historical data. For each grid and each of the four directions, one scatter plot is generated from historical data. Four of these plots are shown in the figure, with their grid locations marked by red circles and directions indicated by arrows in (\textbf{B}). Data points in (\textbf{A}) correspond to the ten-minute average occupancy ($x$-value) and flux ($y$-value). Green dots and blue dots are two selected groups of data points whose centers (red triangles) determines the shape of the fundamental diagram (black solid curve). The model parameters $\mathcal{V}_f, \mathcal{V}_s, \mathcal{N}_c, \mathcal{N}_m$ can then be derived from the fundamental diagram according to the procedure detailed in Sec.~\ref{subsec_driving}. The heat-map in right-lower panel shows the estimated $\mathcal{V}_f$ (only the maximum one out of the $\mathcal{V}_f$-values in four directions is shown). Note that the estimated speeds on oblique roads tend to be larger than those on horizontal and vertical roads because of the zigzagged representation in the gridded road system.} 
 \label{figVN}
\end{figure}

To address the first issue, we try to learn a $V$-$N$ relation from the historical data. In particular, we compute the historical $V_i$ and $N_i$ at every ten-minute time windows (non-overlapping) according to Eqs.~\eqref{eq_averagespeed} and~\eqref{eq_occupancy}. For each grid, these historical values generates a figure like the one showing in Fig.~\ref{figVN}\textbf{A}. In the figure the $x$-axis is $N_i$ and the $y$-axis is $N_{i} V_{i}$, i.~e., the average flux. The pattern looks similar to the classical \emph{fundamental diagram}~\citep{helbing2001traffic}: for small occupancy, the flux increases linearly with road occupancy; When the occupancy exceeds a threshold (vertical dashed line), traffic congestion may occur. A small difference is that, when the occupancy is small, the speed can sometimes be very low. This is probably due to the fluctuations of the ratio of taxis among the whole traffic (so that even when the occupancy of the taxi alone is small, the total occupancy for all vehicles may be big).

For each road link, we fit a $V$-$N$ relation (the thick-black curves shown in Fig.~\ref{figVN}\textbf{A}) from the data points. First we determine the two points \textbf{P} and \textbf{Q} (represented by red triangles) in the figure, which corresponds to the center of the green and blue points, respectively. Currently we adopt a naive way of selecting the green and blue points: first we rule out the top 5\% data points with the largest flux as outliers. Then we select the left-most 20\% from the top-most 20\% data points as the green ones. Lastly we select all the data points on the right-hand-side of the green dots as the blue ones. Then we assume that $V=\mathcal{V}_f$ is constant when $N \le \mathcal{N}_c$ and takes the form $V = a/(N - b)$ when $N > \mathcal{N}_c$, where $a$ and $b$ are two constants whose value are to be determined. To fit this type of $V$-$N$ function we need four parameters, namely $\mathcal{V}_{f}$, $\mathcal{V}_{s}$, $\mathcal{N}_{c}$, $\mathcal{N}_{m}$. $\mathcal{V}_{f}$ is the \emph{free-flow speed} of the road segment, which equals to the slope of the line-segment \textbf{OP}. $\mathcal{N}_{c}$ is the \emph{effective capacity} for taxis on the road segment, corresponds to the $x$-value of point \textbf{P}. The other two parameters, $\mathcal{V}_{s}$ and $\mathcal{N}_{m}$ corresponds to the slope of line-segment \textbf{OQ} and the $x$-value of \textbf{Q}, respectively. In terms of these four parameters, the function $V(N)$ can be written as
\begin{equation}
 V(N) = \begin{cases}
    \mathcal{V}_f,   &  \text{if $N \le \mathcal{N}_c$,} \\
    \displaystyle  \frac{\mathcal{N}_f\mathcal{V}_s(\mathcal{N}_m - \mathcal{N}_c)}{(\mathcal{V}_f - \mathcal{V}_s)N - 
    (\mathcal{V}_f\mathcal{N}_c - \mathcal{V}_s\mathcal{N}_m)},   &  \text{if $N > \mathcal{N}_c$.}
 \end{cases} 
 \label{eq_VN}
\end{equation}
From this $V(N)$ function we can predict a target speed for a given occupancy.

$\mathcal{V}_{f}$, $\mathcal{V}_{s}$, $\mathcal{N}_{c}$, $\mathcal{N}_{m}$ estimated this way give a quantitative description of the intrinsic properties of the road network (Fig.~\ref{figVN}\textbf{B}). Since there are four directions in each grid, the total number of parameters we need to learn from the data is $256\times 256\times 4\times 4=1,048,576$. There are many grids that are seldom visited by cars, and we simply assign a default value to them because these regions do not affect the overall traffic much (we use $\mathcal{V}_{f}=20$km/h, $\mathcal{V}_{s}=5$km/h, $\mathcal{N}_{c}=1$, $\mathcal{N}_{m}=0.5$). All these parameters are pre-computed and stored.

Now we address the second issue, which is to update vehicle positions when the speed $V_i$ is given. As we mentioned ealier, from the historical data we know when and where a car enters into the grids. We also know the route it will take. Denote $l^j(t)$ as the position of the $j$-th car on its pre-determined route at time $t$. $l^j = 0$ corresponds to the starting point and $l^j$ is increased by 100m every time it passes a grid. During the simulation, at each time step we update $l^j$ by
\begin{equation*}
  l^j(t+dt) = l^j(t) + v^j(t)dt,
\end{equation*}
where $dt$ is the simulation time step size (we choose $dt$ to be one minute), and $v^j(t)$ is the speed of the car. $v^j(t)$ is determined by
\begin{equation}
 v^j(t) = \lambda V_{current} + (1-\lambda) V_{next} + \eta.
 \label{eq_speed}
\end{equation}
Here $V_{current}$ and $V_{next}$ is the average speed of the grid where the car is currently at and the next grid that the car is heading to, respectively. $\lambda$ is a constant controlling the \emph{look-ahead} effect~\citep{sopasakis2006stochastic}, i.~e., the impact of the speed of the upstream cars on the downstream cars, and $\eta$ is a random noise. Note that a car may travel multiple grids within one time step. Once the car arrives its destination, it is removed from the system. 

\subsection{Traffic flow simulator}
\begin{figure}
 \centering
 \includegraphics[width=.6\textwidth]{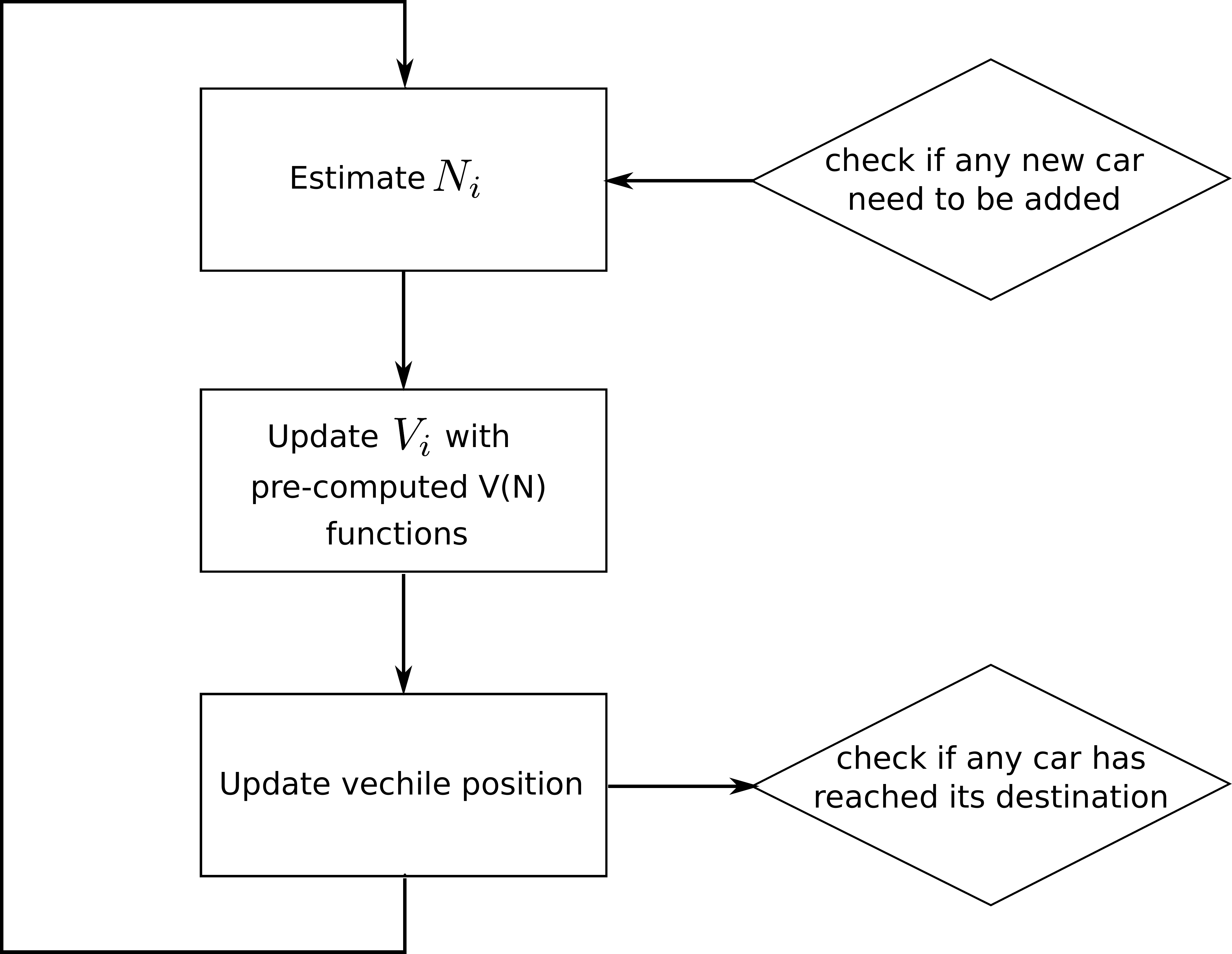}
 \caption{Workflow of the traffic simulator (see main text for details). }
 \label{fig_workflow}
\end{figure}

Now that we have all the components needed to build our traffic flow model. As illustrated in Fig.~\ref{fig_workflow}, at every time step, we add new cars with pre-determined routes to the system. Then we compute the occupancy $N_i$ based on the position of every vehicles on grids. From $N_{i}$ we predict a target driving speed $V_{i}$ using Eq.~\eqref{eq_VN}. This target speed will be used to adjust the current speed on each grid (see below). Once the new speed is known, we update the position of every cars using Eq.~\eqref{eq_speed}. During the simulation, when a car reaches its destination, it is removed from the system. The above process can be repeated to generate the simulated traffic flow dynamics. 

In the following we discuss some miscellaneous points that are important for the robustness and accuracy of our traffic simulator.

\subsubsection{Parameter turning}
\label{subsec_paratune}
The parameters $\mathcal{V}_{f}$, $\mathcal{V}_{s}$, $\mathcal{N}_{c}$, $\mathcal{N}_{m}$ at each grid learned directly from the data provide a good starting point for the model. However, they may subject to errors (or certain type of bias) during the fitting processes and hence are not optimal. As we shall see later, the simulation results are quite sensitive to the choice of parameters. In order to improve the accuracy of the model, we need to fine tune the parameters. To do so, we simply iteratively adjust $\mathcal{V}_f$, $\mathcal{V}_s$, $\mathcal{N}_c$ and $\mathcal{N}_m$ in the following way. First, we define $V_{simulate}^{regular}$ and $V_{real}^{regular}$ to be the \emph{average} of speed during regular hours (from 12:00 to 15:00) obtained from the simulated and historical data at each grid, respectively. If $V_{simulate}^{regular} > V_{real}^{regular}$ we reduce $\mathcal{V}_f$ at this grid by a small amount. Vice versa, if $V_{simulate}^{regular} < V_{real}^{regular}$ then $\mathcal{V}_f$ is increased by a small amount. After many times of iterations, $\mathcal{V}_f$ should approximate the mean speed during the selected regular hour. Next we define $V_{simulate}^{rush}$ and $V_{real}^{rush}$ to be the \emph{minimum} speed recorded during the morning rush hour (7:00-9:00) at each grid. Similarly, if $V_{simulate}^{rush} > V_{real}^{rush}$ we reduce $\mathcal{N}_c$ at this grid by a small amount, and if $V_{simulate}^{rush} < V_{real}^{rush}$ we increase $\mathcal{N}_c$. When changing $\mathcal{N}_c$, we also change $\mathcal{N}_m$ accordingly to keep the ratio $\mathcal{N}_c/\mathcal{N}_m$ unchanged. As we shall see in the numerical results, after using this simply intuitive way to optimize the parameters, the accuracy of the model is greatly improved. We will investigate other parameter tuning methods in future work.
 
\subsubsection{Incorporating real time traffic data}
\label{subsec_realtime}
As the simulation runs over time, errors will accumulate. Using real time traffic data to adjust the simulated traffic state can improve the prediction accuracy. In numerical weather prediction, this kind of procedure is called data assimilation and can be done using methods like Kalman filtering~\citep{kalman1960new}. In our case, we correct the simulated speed using recently estimated speed at each grid (for example, before predicting 10 o'clock traffic at 9:00, we would use the historical speed at 9:00 to update the current speed). For simplicity we assume the noise in the estimated speed is neglectable compared with the noise in the simulation, so we simply replace the simulated speed as soon as we have a set of newly estimated speed. Updating occupancy, however, is more tricky and is not implemented here.
 
\subsubsection{Speed relaxation}
\label{subsec_relax}
To make the model more realistic, we try to avoid sudden jumps in speed by introducing relaxation when updating the speed. Assuming $V_{target}$ is the speed computed according to Eq.~\eqref{eq_averagespeed}, and $V_{current}$ is the current speed which needs to be updated. The predicted speed is given by
 \begin{equation}
  V_{predict} = V_{current} + \omega (V_{target} - V_{current}).
  \label{eq_predict}
 \end{equation}
 Here $0 < \omega < 1$ is a relaxation coefficient (we choose $\omega = 0.5$ as it gives the most reasonable output).

\subsubsection{Private cars}
\label{subsec_private}
\begin{figure}
 \centering
 \includegraphics[width=.6\textwidth]{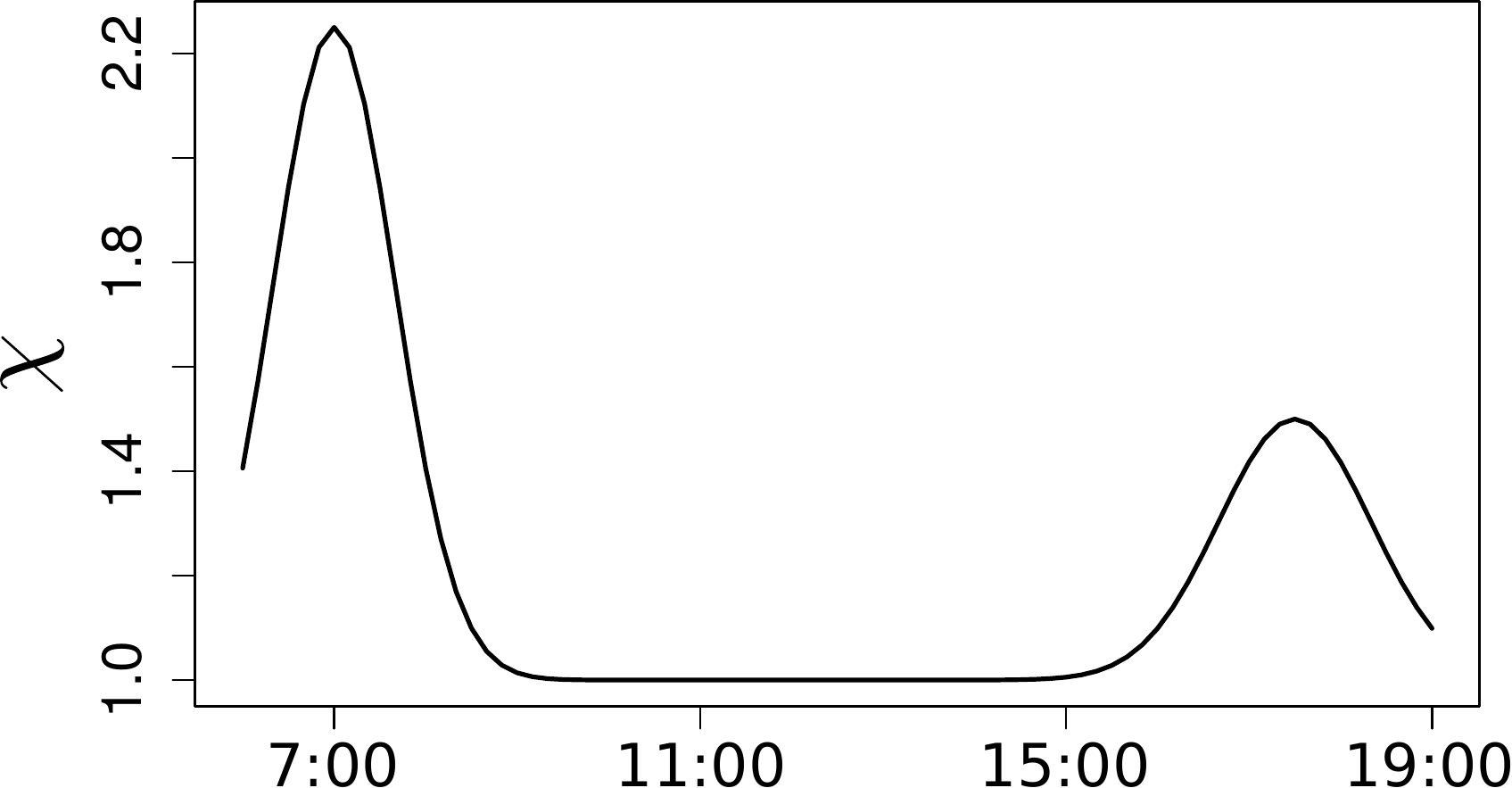}
 \caption{The time-varying capacity coefficient $\chi$ that mimics the rush hour traffic.}
 \label{fig_chi}
\end{figure}

Limited by the data, our model does not explicitly include private cars, which are the most dominant composition of urban traffic. In our model, the effect of private cars is absorbed into the function $V(N)$. One difficulty we encountered is that the proportion of private cars in the traffic is changing over time. In particular, the amount of private cars on road during the morning and afternoon rush hours will drastically increase. Currently we model this effect by multiply $N_i$ on every grids by a time-varying capacity coefficient $\chi(t)$. As shown in Fig.~\ref{fig_chi}, the value of $\chi$ stays 1 at regular hours and reach its peaks at 7:00 and 17:30 during rush hours (two Gaussian functions are used). The choice of $\chi(t)$ is also fitted from data.

\section{Results}
\label{sec_result}
To demonstrate the potential of our model in traffic forecasting in large cities, we use it to predict road speed one hour into the feature in Beijing. We choose a day (2013-03-05, Tuesday) and compute the average speed in every 10 minutes from the historical data. The historical speed information is used for two purposes: one is to serve as the ground-truth for comparison with the model predicted results, and the other is to be used as the ``real time'' input data for runtime speed correction (see Sec.~\ref{subsec_realtime}). We construct two sets of O-D and route information, one come from the day to be simulated, and the other comes from another weekday (2013-03-06). We use this alternative route-set (referred as route-set 2 in the following) to show how different route choice would affect the simulation results. Additionally, in order to demonstrate how sensitive the simulation results depend on model parameters, we consider two parameter sets. One is directly learned from the historical data (and thus may be noisy), and the other is the result of parameter tuning (see Sec.~\ref{subsec_paratune}). The former is referred as parameter-set 2 in the following. With different route set and parameter set, we run the simulation and predict the speed one hour into the future during time period 6:20 to 18:00 of the selected day (results obtained on different days are similar).\\

\begin{figure}
 \centering
 \includegraphics[width=.68\textwidth]{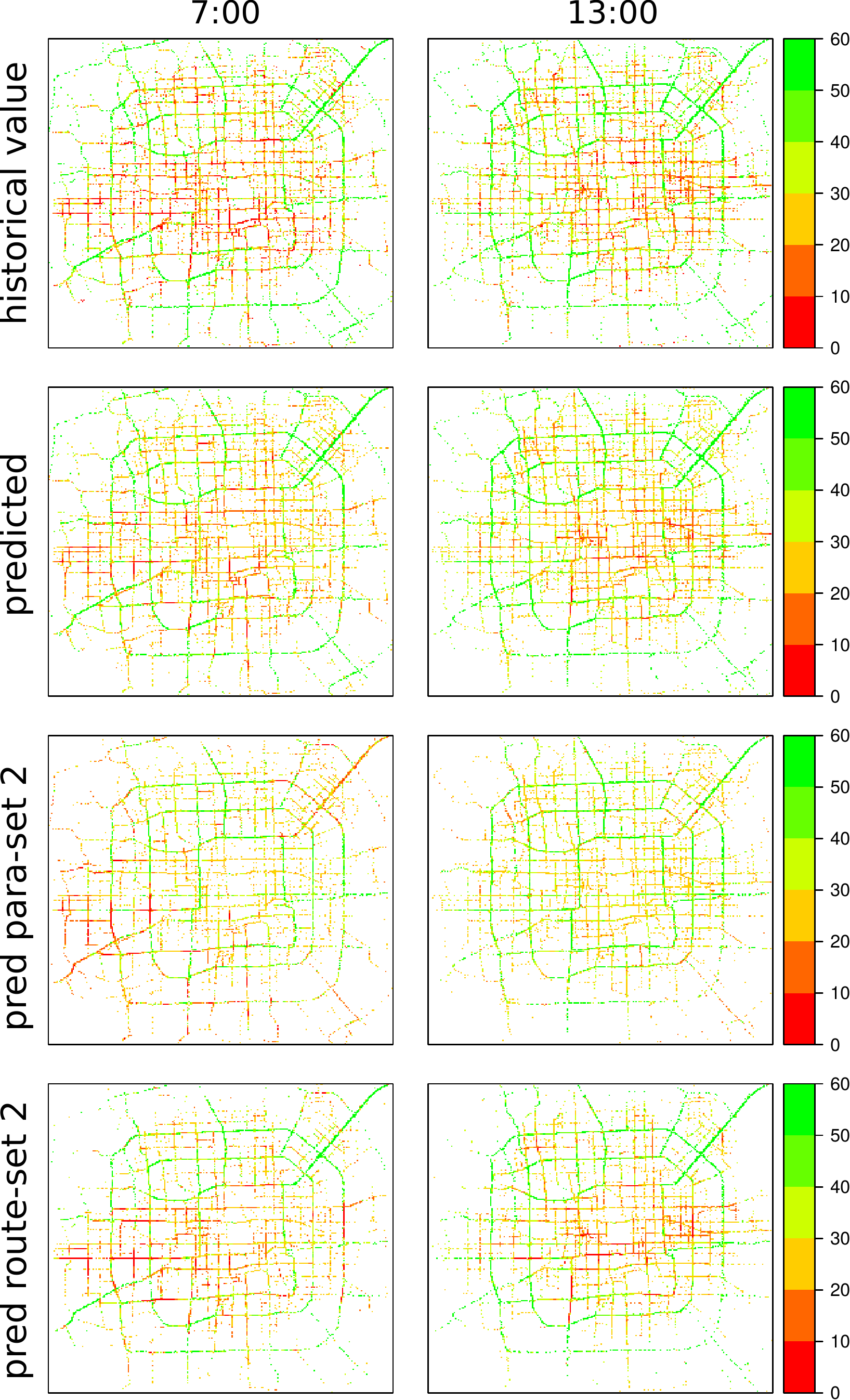}
 \caption{Historical and predicted speed of the entire simulation domain. The left and right column shows the speed at time 7:00 and 13:00, respectively. The speed at each grid is a weighted sum of speeds in the right and up directions. From top to bottom the rows show the speed of the historical value (1st row), one hour prediction of the model (2nd row), one hour prediction of the model using parameter-set 2 (3rd row), and one hour prediction of the model using route-set 2 (4th row). }
 \label{fig_global}
\end{figure}

\noindent\textbf{Global traffic patterns}\\
First we test if our traffic simulator can generate traffic state that is consistent with that observed in the historical data on a city-wide scale. Here \emph{consistent} is in the sense that the predicted congested locations agree with the historical data in a qualitative manner. In Fig.~\ref{fig_global} we plot the speed predicted for 7:00 and 13:00. In each plot we show the combined right and up speed. The combination is done via a weighted sum, with weights equal to the volumes in the corresponding directions. We can see that, the model predicted traffic speed (second row) is in general consistent with the historical data. On the other hand, there appears to be a large discrepancy when using parameter-set 2 (third row), indicating that the model results is sensitive to the choice of parameters. For route-set 2, the overall pattern remains similar with that of the route-set 1, suggesting that the simulation results are not very sensitive to travel demands. This is an important point which makes it possible for us to replace the exact route-set with simulated ones in future (this needs to be done in practical traffic prediction because we do not have the \emph{a priori} knowledge of the travel demands).\\

\begin{figure}
 \centering
 \includegraphics[width=.8\textwidth]{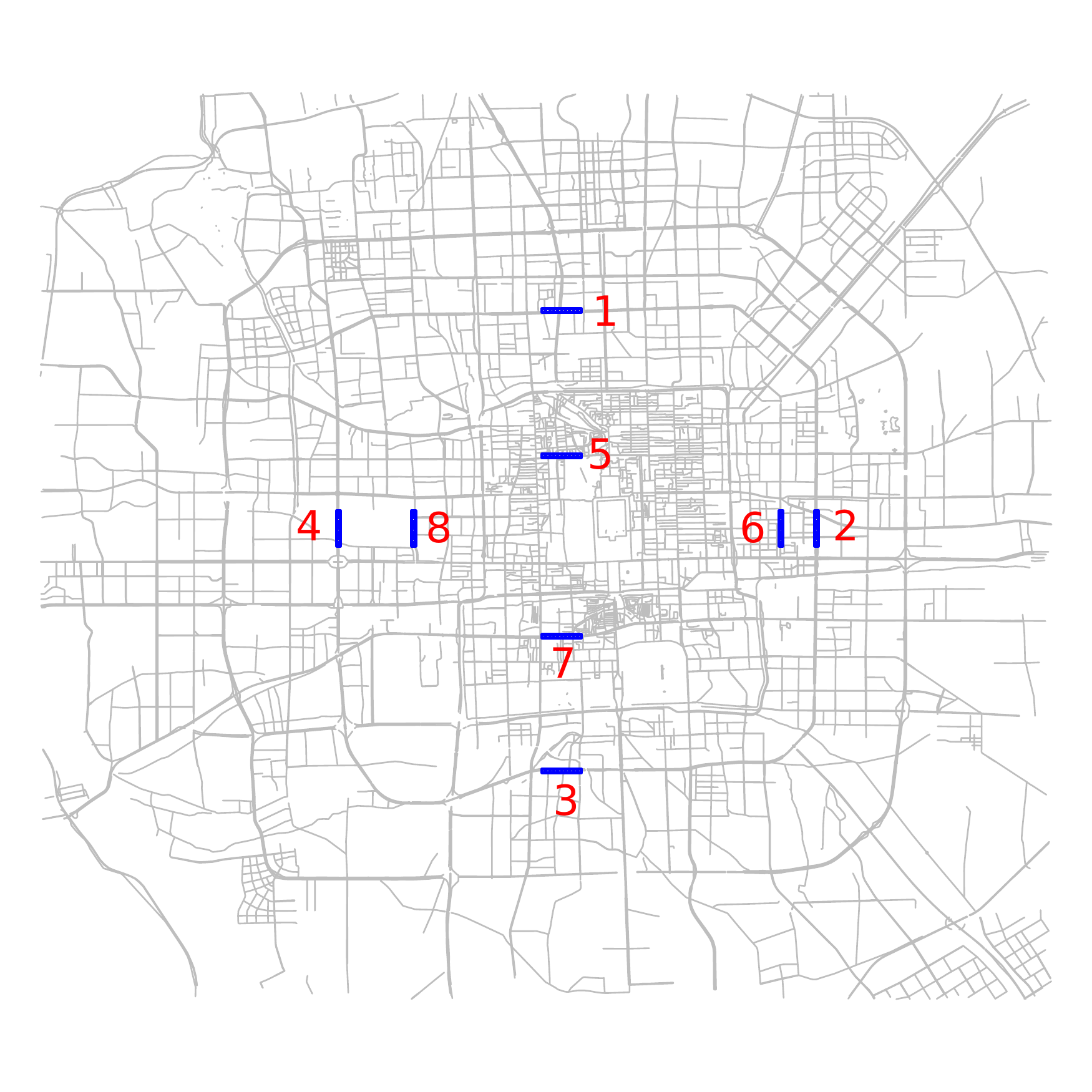}
 \caption{Selected road segments for quantitative comparison. Road segments $\{1, 3, 5, 7\}$ are heading to the right direction, $\{2, 4, 6, 8\}$ are heading to the up direction. Road segments $\{1, 2, 3, 4\}$ are part of arterial roads, and $\{2, 4, 6, 8\}$ are part of local roads. Each road segment starts either from the 128-th row or 128-th column on the $256\times 256$ simulation grids and runs for 10 grids following its direction.}
 \label{fig_roadsegment}
\end{figure}

\noindent\textbf{Local road segment speed}\\
Next we take a closer look at the predicted traffic speed on road segments. To do so, we select a number of road segments, each of which is made of ten connected grids (total length is 1km). We select eight road segments as shown in Fig.~\ref{fig_roadsegment}. Four of them are part of arterial roads, while the other four are part of local roads. The speed of a road segment is defined as the average speed of its compositing grids. To avoid bias in selection we let the starting grid of all road segments to be either 128-th row or 128-th column.

\begin{figure}
 \centering
 \includegraphics[width=.8\textwidth]{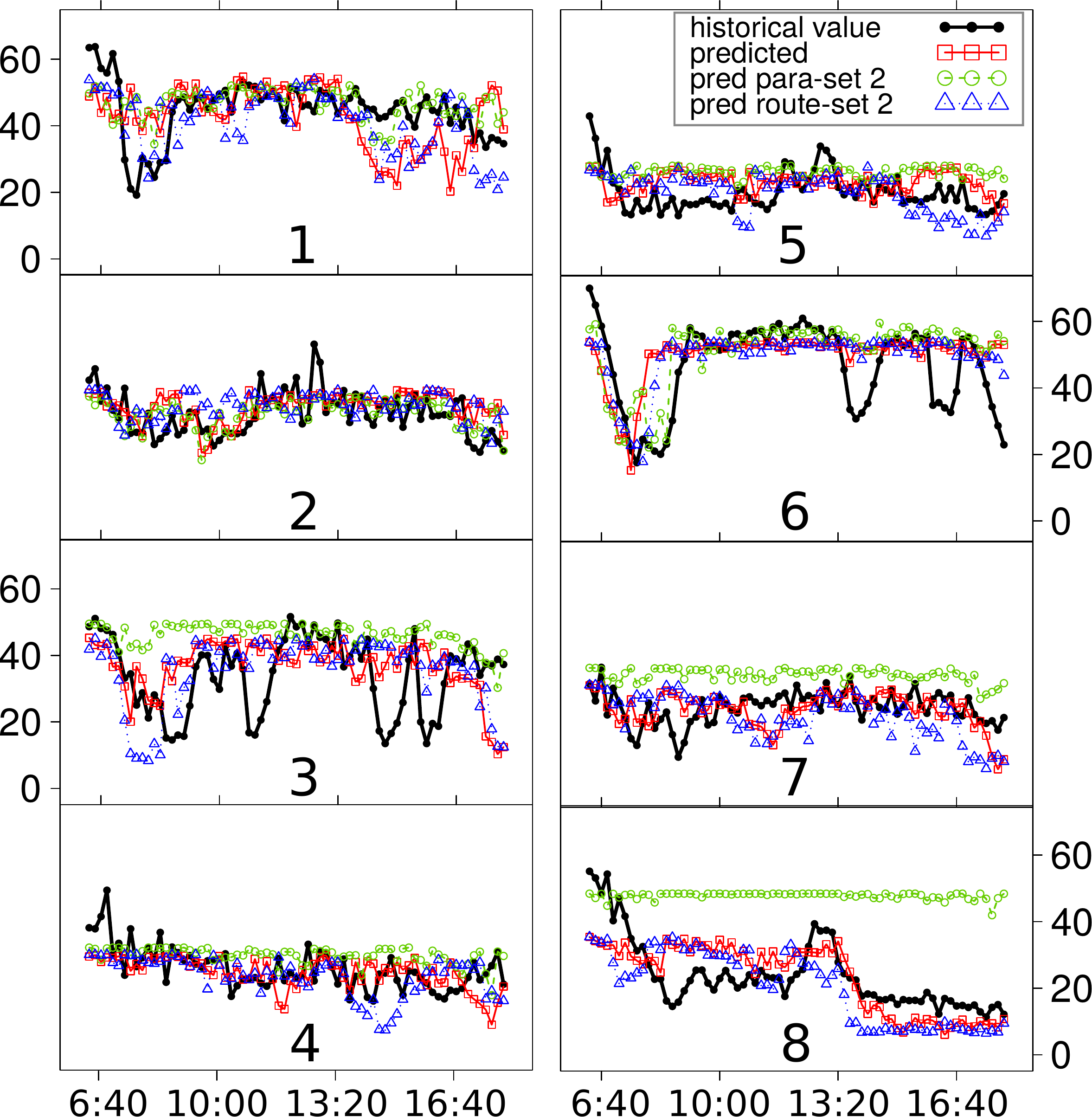}
 \caption{Historical and predicted speeds on eight selected road segments. The simulation time ($x$-axis) spans from 6:20 to 18:00. $y$-axis corresponds to the speed (unit: km/h). The number at the bottom of each panel corresponds to the road segment number shown in Fig.~\ref{fig_roadsegment}. Thick black curve --- historical value; red square curve --- one hour model prediction; green circle curve --- one hour model prediction with parameter-set 2; blue triangle curve --- one hour model prediction with route-set 2.}
 \label{fig_local}
\end{figure}

Fig.~\ref{fig_local} shows the predicted road segment speed using different implementations. The model predicted speed (red square curve) agrees well with the real value on most of the selected roads, but there are significant discrepancy for road segment 3 and 6. In these two cases, there are sudden drops in speed during regular traffic hours. Our current model fails to catch these irregular behaviors in urban traffic, probably because of insufficient data. If using parameter-set 2, the predicted speed (green circle curve) is quite different from the real one, and in road segments 7 and 8 there is even severely deviation in the average of the speed. This explains why parameter tuning is important. On the other hand, using route-set 2 does not change the results much.

\begin{table}
\centering
\begin{tabular}{|c||c|c|}
\hline
 & RMSE & Accuracy \\
\hline
model prediction & \textbf{9.55} & \textbf{0.72}\\
\hline
extrapolation & 11.77 & 0.70 \\
\hline
parameter-set 2 & 13.15 & 0.55 \\
\hline
route-set 2 & 9.60 & 0.70 \\
\hline
\end{tabular}
\caption{One hour prediction error in road segments speed. For each method, both RMSE (the smaller the better) and Accuracy (the larger the better) are computed. See main text for detailed information.}
\label{tab_traveltime}
\end{table}

To further quantify the prediction accuracy, we compute the Rooted Mean Square Error (RMSE) and Accuracy (one minus the Mean Absolute Percentage Error, or MAPE): 
\begin{equation}
 \textrm{RMSE} = \sqrt{\frac{\sum_{i=1}^{T} (V_{real} - V_{predict})^2 }{T}}.
\end{equation}
\begin{equation}
 \textrm{Accuracy} = 1 - \textrm{MAPE} = 1 - \sum_{i=1}^{T} \frac{|V_{real} - V_{predict}|}{V_{real}}.
\end{equation}
In both equations the summation is taken for the whole simulation time period and then averaged over all the selected road segments. As a control experiment, we also consider an \emph{extrapolation} method of prediction, which simply uses the current speed as the prediction. In Table~\ref{tab_traveltime}, the RMSE and Accuracy between the model prediction and real value are given. We can see that, the model prediction scores the best result in both RMSE and Accuracy compared with others. Also, when using route-set 2 the prediction accuracy do not degrade much. Note that we do not attempt to compare the current model with other machine-learning based prediction here. In fact, for now the state-of-art machine-learning prediction program can do better (for example, \citep{min2011real} reported an accuracy of 0.84, and~\citep{lv2015traffic} reported an accuracy nearly 0.94 for one hour prediction). We can only conclude that the current model gives a reasonable prediction, but further improvements need to be made.\\

\begin{figure}
 \centering
 \includegraphics[width=.8\textwidth]{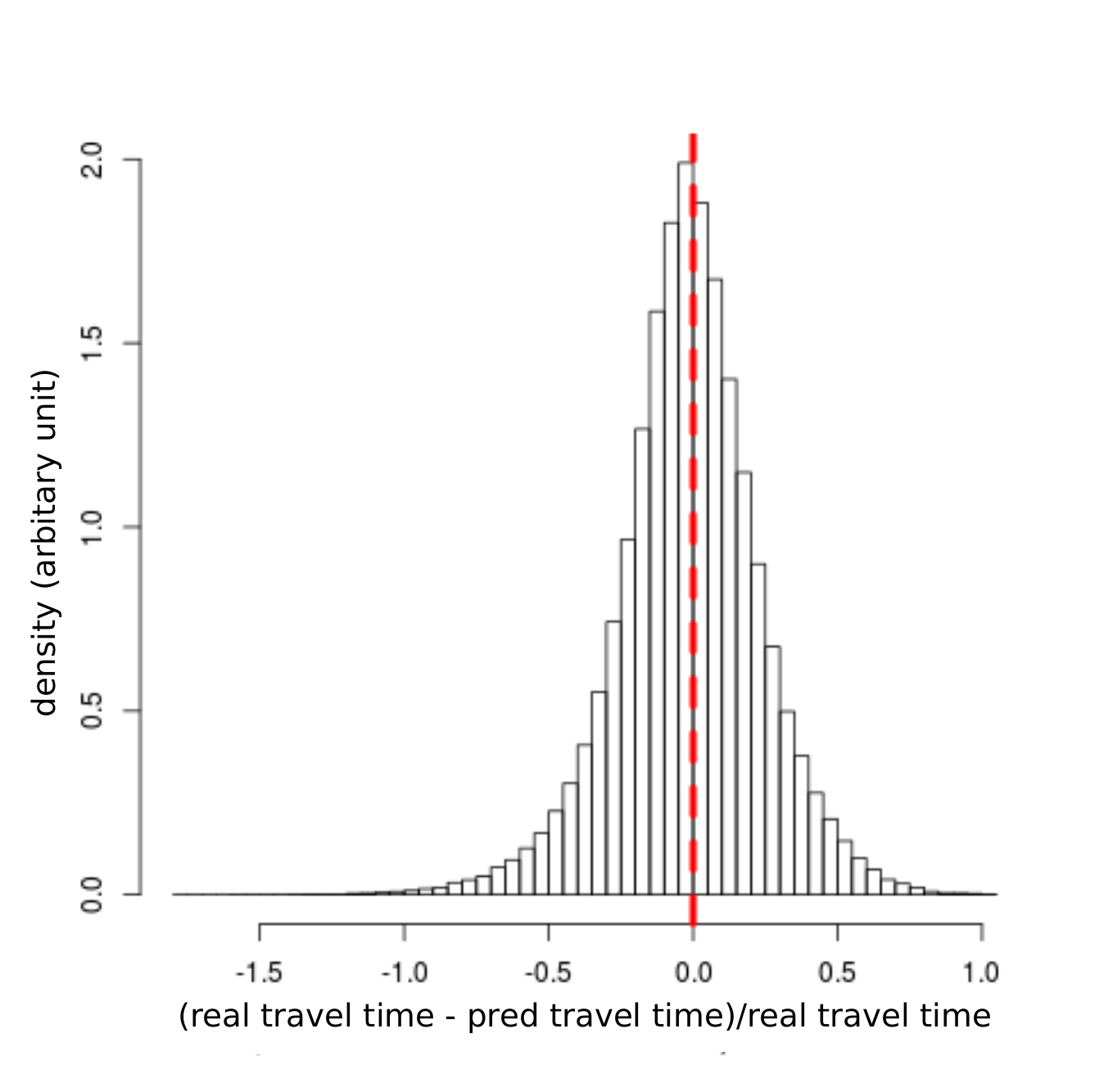}
 \caption{Histogram of the relative error of estimated travel time. \emph{real travel time} corresponds to the recorded time lapse between the destination and origin in the floating car data. \emph{pred travel time} for each trip is estimated based on the one hour ahead predicted traffic state. The red-dashed vertical line separates the positive (under-estimation) and negative (over-estimation) values.}
 \label{fig_traveltime}
\end{figure}

\noindent\textbf{Travel time estimation}\\
One benefit of the model-based method is that the travel time of each trip naturally comes out from it (in machine-learning method travel time estimation is usually completely different than the speed prediction). Fig.~\ref{fig_traveltime} shows the histogram of the relative divergence between the model predicted and historical travel time. The samples that produces this histogram include all the trips whose real trip time falls between 20min$\sim$30min. For most trips, the estimated travel time has an error smaller than 50\%. However, there are a number of trips exhibits large errors. The reasons may due to the model or the data itself (it is possible that some taxi may stop during the trip). While the detailed analysis will be left for future work, the current result shows that the model provide a roughly unbiased travel time estimation.

\section{Conclusions and future work}
\label{sec_discuss}
In this paper we propose a coarse-grained cellular automata model for city traffic forecasting. Building a model that can capture the urban traffic dynamics is a challenge task. We do not aim to faithfully model the real traffic system with fully details. Instead, limited by the computation power and data, we design a model to describe the effective dynamics of the taxis. By leaving many details in the traffic system to be unresolved, the model is simple enough to be implemented on a city-wide scale. It is interesting to mention the similarity between our method and numerical weather prediction. In the later, it is also impractical (due to limited computation power and unpredictable effects) to solve the real physical system in the finest scale. Instead, weather models are build on grids with relatively low spatial resolution, leaving many detailed physical processes in the sub-grid level to be unresolved. As a result, the dynamics on the resolved coarse grid is an effective system, whose model parameters also need to be adjusted using the observation data. 

As part of an on-going work, our model can be considered as a prototype of city traffic forecasting system. We show that our model gives reasonable prediction to the city traffic. Currently we are using only the taxi GPS data. With more data in various form (such as the total traffic volume data, and the traffic accident report data) available in the future, there is ample room for further improvements. Some thoughts are discussed below.

\subsubsection*{Incorporating the information of other vehicles}
The traffic flow of the taxis constitute a relative small portion of the total traffic in Beijing (about 10\% based on rough estimation). To make the problem even more difficult, the ratio of taxis among all the vehicles changes over time and space. For example, in suburbs the taxi ratio tends to be high in the morning, while in urban regions tends to increase in the evening. For now, we only fitted the overall proportional change of taxis by rescaling the capacity everywhere using the same function $\chi(t)$. This is far from ideal. To solve this problem we need more information, say the total traffic volume on a number of representative roads. From them we should be able to estimate the proportion of taxis on these road, and then extrapolate the result to other roads that do not have the data. If we can correctly estimate the ratio of taxis on every roads, the accuracy of the model will likely to be greatly improved.

\subsubsection*{Gridded road system vs. road network}
The real road system is a complex network. It contains many subtle details that may affect the traffic flow, such as traffic light, turnability and the splitting or merging of lanes. For the coarse-grained grid system we use here, these details are neglected. The benefit of this treatment is that the model is easy to setup: as long as the GPS data is available, the model can be easily deployed in other cities without the need to construct a detailed road network first. However, it does have some downsides. For example, we may split the traffic flow of one road into two grids, or merge the traffic flow of two separate roads into one grid. With slight modification, our method can be applied to a complex road network (we can also learn a fundamental diagram for each road link and the model framework is still the same). It would be interesting to see how the model performs on more realistic road networks.

\section*{Acknowledgements}
This work is supported by Beijing Municipal Science and Technology Program (grant Z151100002115040), National Science Foundation of China (grant 11301294 and 11671415) and Tsinghua University Initiative Scientific Research Program (grant 20151080424). The authors thanks Xuesong Zhou for useful discussions.

\bibliography{ref}
\bibliographystyle{plainnat}

\end{document}